\documentstyle[11pt,aaspp4,flushrt]{article}

\begin{document}

\title{DDO Photometry of M71: Carbon and Nitrogen Patterns Among Evolving Giants}

\author{Michael M. Briley}
\affil{Department of Physics and Astronomy,\\University of Wisconsin Oshkosh,
Oshkosh, Wisconsin 54901}

\author{Graeme H. Smith}
\affil{University of California Observatories/Lick Observatory, University of
California,\\ Santa Cruz, California 95064}

\author{C. F. Claver}
\affil{Kitt Peak National Observatory, National Optical Astronomy Observatories,
\\P.O. Box 26732, Tucson, Arizona 85726}

\begin{abstract}
We present $V$, $B-V,$ and DDO $C(41-42)$ and $C(42-45)$ photometry for a sample
of 75 red giants down to $M_V = +2$ in the relatively metal-rich Galactic globular
cluster M71. The $C(41-42)$ colors reveal a bimodal distribution of CN band
strengths generally anticorrelated with CH band strength as measured by the
$C(42-45)$ color. Both DDO colors agree well with those found in 47 Tucanae
-- a nearby globular cluster of similar metallicity -- and suggest nearly identical C and N
abundance patterns among the giants of both clusters. A comparison with synthetic
DDO colors demonstrates that little change in surface C or N abundance is required 
to match the colors of the M71 giants over the entire luminosity range
observed. Apparently like 47 Tuc (a cluster of much greater mass
and central concentration), M71 exhibits an abundance pattern which
cannot be solely the result of internal mixing.

\end{abstract}


\section{Introduction}

Over the past three decades, beginning with the observations of Osborn (1971), it
has become evident that the stars within Galactic globular clusters are not
chemically homogeneous populations. Significant star-to-star variations in C, N,
and O, and often Na, Mg, and Al have been reported among the red giants in most
Galactic globular clusters for which these elements have been investigated
in-depth. Reviews of this subject can be found in Kraft (1994), Da Costa (1998),
and Cannon et al. (1998). The observations to date indicate that some mechanism or
mechanisms have acted to modify the light element composition of the cluster stars
while leaving heavier elements (Ca, Fe, etc.) relatively untouched. Proton-capture 
reactions involving C/N/O/Na/Mg/Al at the temperatures of the C$\rightarrow$N and
O$\rightarrow$N cycles appear to be implicated (Denissenkov \& Denissenkova 1990;
Langer, Hoffman, \& Sneden 1993; Cavallo, Sweigart, \& Bell 1996, 1998).
What remains unclear is the location of these reactions, how they lead to the
inhomogeneities in the cluster stars, and what this implies about the early
cluster environment and the structures of low-mass stars.

One relevant nucleosynthesis site lies just above the hydrogen-burning shell of
an evolving cluster giant where it has been shown theoretically that conditions are
sufficient not only for CN and ON-cycle reactions, but for proton captures on Ne
and Mg (yielding Na and Al) as well (see, for example, the above references as
well as Langer \& Hoffman 1995). The Ne-Na and Mg-Al cycles that can take place at
the temperatures of the  CNO-burning shell are described in Cavallo et al. (1996).
Should this material be accessible to some form of deep dredge-up, one could
expect changes in the surface abundances of C (decreasing with time), N
(increasing), O (decreasing), as well as possible changes in Na, Al, and Mg (e.g.,
Shetrone 1996a,b). Mixing down to the outer parts of the CNO-burning shell is not
predicted by standard models of low-mass giants  (Iben 1964), although several
mechanisms have been explored (Sweigart \& Mengel 1979; Charbonnel, Brown, \&
Wallerstein 1998; Fujimoto, Aikawa, \& Kato 1999; Denissenkov \& Tout 2000).

While observations of low $^{12}$C/$^{13}$C ratios (Brown \& Wallerstein 1989;
Bell, Briley, \& Smith 1990; Brown, Wallerstein, \& Oke 1991; Suntzeff \& Smith
1991; Briley et al. 1997) and various abundance patterns such as O versus Na and
Mg versus Al anticorrelations (e.g., Sneden et al. 1997; Kraft et al.  1997; Ivans
et al. 1999; Cavallo \& Nagar 2000) may be consistent with deep  mixing, a direct
test of whether or not such mixing is taking place is to look for composition
changes as a function of evolution (luminosity) on the red giant branch (RGB).
Indeed, C abundances that decrease by a factor of ten with increasing luminosity have been
observed among the evolving giants of the more metal-poor ([Fe/H] 
$\approx -2$\footnote{For the purpose of discussion, we have loosely characterized
the metallicities of globular clusters as ``poor,'' ``intermediate,'' or ``high.''
See Harris 1996 for a compilation of abundance data from the literature.})
clusters M92 (Bell, Dickens, \& Gustafsson 1979; Carbon et al. 1982; Langer et al.
1986; Bellman et al. 2001), M15 (Trefzger et al. 1983), and NGC 6397 (Briley et al. 1990), as
well as increasing Na abundances with luminosity in M13 (Pilachowski et al. 1996). A
similar trend was also found among the intermediate metallicity ([Fe/H] $\approx -1.5$)
clusters M3 and M13 by Suntzeff (1981, 1989), and for NGC
6752 by Suntzeff \& Smith (1991). Results such as these provide almost irrefutable
evidence that at least some cluster stars are somehow mixing newly synthesized
material to their surface. However the picture for higher metallicity globular
clusters ([Fe/H] $\approx -0.8$) is not so clear. Trends in [C/Fe] versus [Fe/H] among
upper giant branch stars measured by Bell, Dickens, \& Gustafsson (1979), Dickens, Bell, \&
Gustafsson (1979),  and Bell \& Dickens (1980)  show that the C depletion achieved by
evolved stars decreases with increasing metallicity. The most metal-rich object in
these three studies was 47 Tuc, and Briley (1997) confirmed from analysis of DDO
photometry that little change is seen in C and N abundances over the upper 4
magnitudes of the red giant branch of this cluster. By contrast, in the metal-poor
cluster M92 a change in [C/Fe] of almost 1 dex is observed above $M_V = +2$ (Langer
et al. 1986; Bellman et al. 2001). Such observations imply that evolutionary
processes are altering CNO abundances during RGB evolution in a manner
that is sensitive to overall metallicity.

Other observations indicate that processes interior to the present red giants
cannot be the entire explanation for abundance anomalies in globular clusters.
Studies of CN and CH band strengths of main-sequence turn-off stars in the higher 
metallicity clusters 47 Tuc (Hesser 1978; Hesser \& Bell 1980; Bell, Hesser, \& 
Cannon 1983; Briley et~al. 1991, 1994; Cannon et~al. 1998),  and M71 (Cohen 1999, Briley \& Cohen
2001), as well as a small number of main-sequence stars in the intermediate metallicity
cluster NGC 6752 (Suntzeff \& Smith 1991), have revealed the presence of anticorrelated C and
N abundance inhomogeneities very similar to those found among the evolved giants in these
clusters. Briley et al. (1996) also demonstrated substantial Na differences,
correlated with CN, among three pairs of 47 Tuc turn-off stars. If, as is widely
assumed, these low-mass low-luminosity stars are incapable of appreciable CNO
nucleosynthesis, then it is unlikely that the site of the light-element
proton-capture events needed to produce their CN enhancements lies within these stars.

A number of scenarios have been devised to account for light element inhomogeneities
among such faint stars, each with strengths and weaknesses, many of which are well
summarized by Cannon et al. (1998) and Bell et al. (1981). Of particular interest
to the present work is a sequence of events whereby the present cluster stars
become ``polluted'' with material ejected from intermediate-mass ($\approx 5
M_\odot$) asymptotic giant branch (AGB) stars at $\approx 10^8$ years 
after cluster formation (D'Antona, Gratton, \& Chieffi 1983; Ventura et al. 2001).
The exact specifics of this process depend on a variety of factors; initial mass
function, AGB mass-loss rates, central star density, velocity dispersion, etc. A
critical feature of this scenario is the amount of ejecta which must be assimilated per
star to produce the inhomogeneities observed today, i.e., did CN-strong
main-sequence stars accrete only small amounts of material that mixed
exclusively within their outer convective envelopes, or were larger quantities of
enriched material (i.e., a significant fraction of the star's total mass) mixed
throughout the bulk of their interior? The mass of the  convective envelope in
low-mass cluster stars changes from nearly zero on the  main sequence to more than
70\% of the total stellar mass at the end of first  dredge-up. Thus, if the
accreted material only existed within the surface convection zone during the
main-sequence phase, one should expect  to see a dilution of the inhomogeneities
as the convection zone deepens during RGB evolution. Once
again the behavior of C and N abundances as a function of luminosity on the RGB
can play a role in constraining alternative origins for globular cluster
abundance inhomogeneities.

We present in this paper observations of DDO $C(41-42)$ and $C(42-45)$ colors for a
sample of 75 red giants of M71 spanning luminosities from the giant branch tip to
$M_V \approx +2$ in order to study the luminosity-dependence of [C/Fe] and [N/Fe] 
abundance inhomogeneities in a high-metallicity ([Fe/H] $\sim -0.8$) globular cluster.
Using synthetic colors, we examine whether any changes in
composition with evolution can be detected, due either to mixing or dilution and
compare these results to similar observations of 47 Tuc (a cluster of comparable
metallicity and substantially greater mass).

It was known from the DDO photometry of Hesser, Hartwick, \& McClure (1977) that
M71 contains stars on the upper RGB having similar luminosities and broad-band
colors but very different $\lambda$4215 CN band strengths.  This was confirmed
spectroscopically by Smith \& Norris (1982a), who also found tentative evidence of
a CN-CH anticorrelation. Penny, Smith, \& Churchill (1992) found a bimodal CN-band
strength distribution and anticorrelated CN and CH inhomogeneities among lower RGB
stars in M71, and a similar pattern was found by Smith \& Penny (1989) among
the red horizontal branch (RHB) stars of this cluster. Similar behavior of the
CN and CH bands are also displayed by both upper and  lower RGB stars  (Norris \&
Freeman 1979; Dickens, Bell, \& Gustafsson 1979; Bell et al. 1983; Norris,
Freeman, \& Da Costa 1984; Smith, Bell, \& Hesser 1989; Briley
1997), RHB stars (Norris \& Freeman 1982), and AGB stars (Norris \& Cottrell
1979) in 47 Tuc.

\section{Observations}

Images of M71 were obtained on the nights of 1991 June 8 and 10 with the KPNO
0.9-m telescope and a TE1K CCD using the KPNO DDO and $BV$ filter sets. The 
1024 $\times$ 1024 detector at f/7.5 resulted in 0.73 arcsec pixels for a 12.5 arcmin
field. Seeing varied from 2.0 to 2.5 arcsec under non-photometric conditions on
both nights. One field, centered on M71, was observed for a total of three 1200s
integrations in the 41 filter, $2\times 1200$s in 42, $1\times 1200$s in 45, and a
300s exposure in both $B$ and $V$.

Reductions followed standard procedures for PSF photometry using DAOPhot within
IRAF\footnote{IRAF is distributed by the National Optical Astronomy Observatories,
which are operated by the Association of Universities for Research in Astronomy,
Inc., under cooperative agreement with the National Science Foundation.}. A total
of 11 bright isolated stars were used in constructing constant PSF functions for
each exposure. Instrumental magnitudes were converted onto the standard DDO system 
(McClure 1976) using seven M71 stars with measured photoelectric colors from
Hesser et al. (1977). For these stars the average difference
between the transformed CCD colors and those of Hesser et al. (1977) was
$-0.001 \pm 0.042$ for $C(42-45)$ and $-0.002 \pm 0.023$ for $C(41-42)$.

Likewise, the $V$ and $B-V$ photometry was calibrated to the
photoelectric/photographic survey of Arp \& Hartwick (1971). Comparing 88
stars in common with Arp \& Hartwick produced transformations having standard
deviations $\sigma$(transformed CCD - literature) of 0.046 for $V$ and  0.054 for
$B-V$. We note that highly accurate $BV$ photometry is not required for present
purposes, and random errors of this size in $B$ and $V$  have little affect on the
results.

The 75 RGB program stars were selected from the proper motion survey of Cudworth
(1985) as having greater than 75\% probability of membership. Their locations in
the M71 color-magnitude diagram of Hodder et al. (1992) are shown in Figure 1
(note that as in Hodder et al., only members in the Cudworth sample with greater
than 50\% probability of membership are plotted for $V < 16.85$).

For program stars with multiple observations in the 41 (3 exposures) and 42
(2 exposures) filters, the deviation from the mean magnitude is plotted 
in Figure 2 as a function of $V$ for each star in each exposure. The average
standard deviation about the mean in the 41 filter is 0.018 mags and 0.015
mags in the 42 filter. Errors of this size are consistent with the $\pm 0.023$
uncertainty in the transformation onto the Hesser et al. (1977) $C(41-42)$ system noted
above. Unfortunately, the single observation in the 45 filter precludes a similar
look at errors in the $C(42-45)$ colors.

\section{Comparison to 47 Tucanae}

The clusters 47 Tucanae and M71 are almost identical in overall metallicity (M71 is 
listed as being 0.03 dex more metal-rich than 47 Tuc, for which [Fe/H] = $-0.76$, 
in the compilation of Harris 1996),
and have similar color-magnitude diagrams  (Frogel, Persson, \& Cohen 1981; Hodder
et al. 1992). It is therefore expected that RGB stars of similar
luminosity in the two clusters are at similar stages of evolution, with similar
masses, temperatures, and gravities. As such, differences in molecular band
strengths between giants of the same luminosity in the two clusters are
expected to correspond to differences in atmospheric composition.

With this in mind, the present $C(42-45)$ and $C(41-42)$ photometry for the RGB
members of M71 is compared with that of 47 Tuc from Norris \& Freeman (1979) and
Briley (1997) in Figure 3. Distance moduli and reddenings adopted  are 
$(m-M)_V = 13.80$ and $E(B-V) = 0.27$ for M71 following Hodder et al. (1992), and
$(m-M)_V = 13.40$, $E(B-V) = 0.04$ for 47 Tuc from Hesser et al. (1987). Reddening of the DDO
colors was assumed to be $E(42-45) = 0.234 E(B-V)$ and $E(41-42) = 0.066 E(B-V)$
(McClure 1973). Figure~3 shows the following results:

1) There is an intrinsic spread in $C(41-42)$ among the M71 RGB stars.  As this
color is sensitive to absorption by CN at 4215 \AA, the variations are interpreted
as star-to-star differences in surface CN abundances. The distribution of
CN band strengths, as indicated by $C(41-42)$, appears to be bimodal down to
$M_V \approx +1.$ The $C(41-42)$ error bars are derived from the multiple
observations in the 41 and 42 filters presented in Figure 2 (the standard deviations in
the 41 and 42 filter magnitudes being added in quadrature).

2) The $C(42-45)$ colors (a measure of G-band strength) among the M71 RGB stars
are also consistent with those of the 47 Tuc giants.
As with 47 Tuc, there appears to be an anticorrelation between the
$C(42-45)$ and $C(41-42)$ colors (i.e., a CN-CH anticorrelation), although
the scatter in $C(42-45)$ for the M71 red giants makes a detailed comparison with
47 Tuc difficult. This scatter is not reflected in the corresponding $C(41-42)$ colors, which
suggests that it may be due to random errors in $C(42-45)$, most likely the result of
the single 1800s exposure in the 45 filter.
Because only one 45 frame was taken, the error bars plotted 
in Fig.~3  represent the 0.042 scatter in the comparison of the 11 stars in common 
with Hesser et al. (1977), as opposed to errors derived from multiple exposures.
Despite the scatter, Figure 3 does suggest that the range and distribution of CH band
strengths among 47 Tuc and M71 red giants are similar.

\section{Comparison with Model Colors}

Given the similarity between the M71 and 47 Tuc color-magnitude diagrams, we use
here the same models employed in an analysis of 47 Tuc DDO colors by
Briley (1997), who also provides complete details of how these models were
generated. In short, a grid of model atmospheres was constructed using the MARCS
program (Gustafsson et al. 1975). From the models, synthetic spectra were
generated with the SSG program (Bell \& Gustafsson 1978) and convolved with $V$
and $K$ filter curves. Temperatures and gravities were adjusted to match the
fiducial $V$ versus $V-K$ diagram of Montegriffo et al. (1995) assuming stellar masses of
$M = 0.8$ $M_\odot$ and a metallicity of [Fe/H] = $-0.80$. The final parameters
of the models are given in Table 1, where we also present synthetic
$B-V$ colors for comparison with the observed M71 color-magnitude diagram in Figure~1.

Briley (1997) calculated synthetic spectra matched to the 47 Tuc giant branch
for two different sets of C and N abundances (assuming [O/Fe] = $+0.45$). The
spectra were then convolved with the DDO filter response curves to generate
synthetic colors appropriate to these abundances. One combination of abundances 
([C/Fe] = 0.0, [N/Fe] = $+0.4$) was found to provide a reasonable match to the
colors of CN-weak giants in 47 Tuc, another combination ([C/Fe] = $-0.3$, [N/Fe] =
$+1.4$) was found to match CN-strong giants.  For comparison, Norris \& Cottrell
(1979) found that the abundance differences between a pair of CN-weak and CN-strong
asymptotic branch giants in 47 Tuc
amounted to $\Delta$[C/Fe] = $-0.2$ and $\Delta$[N/Fe] = $+1.0$.
Similarly, Cannon et al. (1998) derived an average [C/Fe] = $-0.15$, [N/Fe] = $+1.05$ 
and [C/Fe] = $+0.06$, [N/Fe] = $+0.20$ for a large sample of 47 Tuc
CN-strong/weak (respectively) main sequence and RGB stars.

Guided by the similarities between M71 and 47 Tuc in Figure 3, we have repeated
these calculations with a similar set of C and N abundances, also taking into
account the [O/Fe] results of Sneden et al. (1994) which were found to anticorrelate
with CN-band strength as described in Briley et~al. (1997).
We have calculated models for the abundance combination [C/Fe] = 0.0,
[N/Fe] = $+0.4,$ and [O/Fe] = $+0.4$ for the CN-weak stars, and [C/Fe] =
$-0.3$, [N/Fe] = +1.4, [O/Fe]= +0.2 for CN-strong giants. The resulting synthetic 
colors are listed in Table 1 and are plotted against the observed colors of the M71 program
stars (corrected for reddening) in Figure 4.

As with 47 Tuc, the CN-weak M71 stars as a group are reasonably well fit by models
with [C/Fe] = 0.0 and [N/Fe] = $+0.4,$ while the CN-strong stars are roughly consistent
with [C/Fe] = $-0.3$ and [N/Fe] = $+1.4$. The situation is less certain with regard to
CH where the sensitivity of $C(42-45)$ to C abundance is fairly weak, becoming
essentially insensitive near the RGB tip due to decreasing blue flux. As discussed
above, the source of the large scatter in $C(42-45)$ is thought to be random errors
in the photometry.

Given the large decreases in surface C abundance found among evolving giants
in significantly more metal-poor clusters (e.g., M92, M15, and NGC 6397, see
above references), we have also calculated a series of DDO colors assuming
progressive deep mixing which reduces the C abundance
to [C/Fe] = $-1.0$ by the RGB tip. While there is no spectroscopic evidence for 
such extreme C depletions in M71, this admittedly extreme exercise allows
us to explore the roles of C and N abundances in setting the present DDO colors.
The C depletion was assumed to begin between the $M_V = $ +0.47 and +1.36 models, 
in accord with the observation by Fusi Pecci et al. (1990)
of a local peak in the luminosity function at $M_V = +0.79$ (in many models of deep 
dredge-up, this point signals the destruction of an interior molecular weight 
gradient and the onset of mixing),
and continues at a relatively constant rate of decline until [C/Fe] = $-1.0$ by
$M_V = -1.50$. Corresponding increases in [N/Fe] were assumed, holding C+N
constant (as would be expected from the mixing of CN-cycle material into the
atmosphere). We have not taken into account mixing of ON-cycle products.
The input abundances and resulting colors are listed in Table 2 and are plotted in Figure 5.

 From these models it can be seen that the overall effect of progressively mixing 
significant quantities of CN-cycle material into the atmospheres of the CN-strong M71
giants is to reduce, not enhance, CN band strengths and corresponding $C(41-42)$ colors.
This result follows from molecular equilibrium -- C is the minority element in
CN formation among most of these C$\rightarrow$N-processed models due to their 
large N abundances, even prior to the assumed onset of mixing for the CN-strong sequence.
Among the CN-weak giants, there is
little change in $C(41-42)$ color with mixing: the slight increase in CN band
strength with increasing N is quickly offset by the decreasing C abundances. 

In the case of the same fractional decrease in [C/Fe] as is seen in the
most metal-poor clusters, the synthetic DDO colors do not seem to provide a good
representation of  the observed run of $C(41-42)$ colors with $M_V$ for the CN-strong
giants. Nor do these deeply-mixed models provide a good fit to the $C(42-45)$ colors of
47 Tuc giants brighter than $M_V \approx +0.5$. A similar situation was also
demonstrated for 47 Tuc by Smith, Bell, \& Hesser (1989).
It is possible to arrive at a somewhat better fit to the DDO colors by including the mixing
of ON-cycle material (freeing up C for CN formation through reduced CO band strengths).
However, such O$\rightarrow$N processing does not seem to accord with the oxygen abundance analyses of
Sneden et~al. (1994), who observed a lack
of extreme O deficiencies among upper-RGB stars of M71 (certainly by comparison
with what is seen in clusters such as M13, for example, by Kraft et~al. 1992, 1993).
In addition they find that the [O/Fe] range among M71 giants is notably smaller than 
within metal-poorer clusters like M13, M3, and M15. 
Thus, if progressive mixing {\it is} occurring along the M71 giant branch, then the
observed pattern of  DDO colors requires that this process be relatively
modest in terms of altering the [C/Fe] ratio.
Indeed, a mixing of the same {\it amount} of CN-cycle material believed
responsible for the 1 dex drop in [C/Fe] of a metal-poor cluster giant would
be diluted in an M71 giant to the point of being essentially undetectable (a 0.03 dex
drop in [C/Fe] by the RGB tip).

\section{Discussion}

The RGB stars of both M71 and 47 Tuc show similar overall
patterns of CN band strengths (Figure 3). The existence of significant
$C(41-42)$ scatter from the RGB tip to at least 1 mag below the HB (see Figure 3)
demonstrates that the CN inhomogeneities are well in place before most
hypothesized RGB-ascent mixing mechanisms are in operation. Other studies of
considerably less-luminous stars in these two clusters (see for example Briley et
al. 1994; Cannon et al. 1998; and Cohen 1999) have reported significant
star-to-star differences in CN and CH band strengths consistent with those found
among the giants. Indeed, Briley \& Cohen (2001) demonstrate that the same set of
abundances which match the present bright giants (i.e., [C/Fe] = 0.0,
[N/Fe] = +0.4, [O/Fe] = +0.4, and [C/Fe] =$-0.3$, [N/Fe] = +1.4,
[O/Fe] = +0.2), also fit spectra of CN and CH bands in M71 main-sequence stars
down to $M_V \approx +4.5$. Such observations suggest an origin to the C and N
abundance variations which precedes the red giant phase in the evolution of
cluster stars, possibly a primordial or a pollution scenario.

Rather than review the various external-enrichment mechanisms that have
been considered in the literature, we refer to Cannon et al. (1998) for a
detailed discussion of this topic. In the remainder of this section we
comment on a few constraints that the 47 Tuc and M71 results bring to this subject.

1) Identical C and N abundances appear to fit both the present giants and the main
sequence stars of Briley \& Cohen (2001). This requires little change in surface
composition despite the deepening convective envelopes during RGB ascent. This has
several ramifications. Firstly, it appears that the giants in 47 Tuc and M71 do
not experience the type of deep mixing that produces a marked decrease in
[C/Fe] with increasing luminosity on the giant branches of more metal-poor
clusters such as M92, M15, M3 and M13. This accords with the observational finding
by Bell \& Dickens (1980), and the theoretical expectations of Sweigart \& Mengel
(1979) and Cavallo et al. (1998), that deep mixing in globular-cluster giants is
more effective at lower metallicity. The former find that [C/Fe] depletions on the
upper giant branches of globular clusters decrease with increasing [Fe/H].
Sweigart \& Mengel (1979)  and Cavallo et al. (1998) show that as the metallicity
of an upper RGB star is increased the regions of C$\rightarrow$N and
O$\rightarrow$N conversion at the outer edge of the  hydrogen-burning shell
become narrower and increasingly closer to the main regions of energy production where
the hydrogen and helium abundances, and the molecular weight, are substantially
altered. If deep mixing within the radiative zone of red giants is inhibited by
molecular weight gradients, then the ability of deep mixing to access the C-N-O
conversion region would be reduced with increasing metallicity. The nature and
behavior of deep mixing within Population II red giants is coming under
increasingly more theoretical scrutiny, and readers are referred to the papers by
Cavallo et al. (1996, 1998), Cavallo \& Nagar (2000), Charbonnel (1995),
Charbonnel, Brown, \& Wallerstein (1998), Denissenkov \& Weiss (1996), Denissenkov
et al. (1998), Denissenkov \& Tout  (2000), Sweigart (1997), Weiss, Denissenkov,
\& Charbonnel (2000).

2) There is a second inference from the result that C and N  abundances of
bright M71 giants and main-sequence turn-off stars appear to be similar.
If main-sequence cluster dwarfs were ``polluted'' with CNO-processed material 
that had been synthesized elsewhere, suppose
that this material was distributed only within the outer convection zone,
which does not comprise much of the mass of these stars. When the convection
zone deepens on the giant branch, and the externally-processed material
is distributed throughout more of the stellar
interior, a decrease in the size of the C and N surface abundance anomalies
might be expected along the RGB (Cannon et al. 1998). However, the observations
provide little indication that C and N abundances decline
along the RGBs of M71 and 47 Tuc. This suggests that prior to the giant phase,
the externally-processed material in CN-strong stars was well mixed
to at least the maximum depth eventually attained by the
convective envelope, which grows to over
70\% of the total mass of the star in standard models. A pollution scenario
therefore requires that any material added to the surface of a
cluster main-sequence star be distributed quite deeply below the base of the
main-sequence convective zone. Some form of interior mass transport, capable
of accessing much of the mass of a main-sequence star seems necessary in
order to make a pollution scenario work.

The occurrence of interior mixing within globular cluster main-sequence stars 
may not be unreasonable. Observations of the Li abundance among Population I
main sequence stars in open clusters show that the Li abundance appears to
evolve with age (Pinsonneault 1997, Deliyannis et al. 1998, Jones et al. 1999), 
indicative of some type of mass transport within the radiative regions of these
stars that connects the base of the convective region with interior regions
hot enough to burn Li.

3) An alternative to the ``external pollution'' scenario for CN-strong giants is
the ``primordial enrichment'' scenario in which such stars formed from gas that
had been enriched prior to star formation. However the primordial enrichment
scenario also has difficulties.  A considerable problem is explaining why
the C abundance of CN-strong giants is less than that of CN-weak stars. If it
is assumed that the CN-weak stars reflect the ``original'' composition of a
globular cluster, then additional material that was C-poor (and N-rich) must
have been incorporated into the
gas from which CN-strong stars formed. However, accounting for the observed [C/Fe]
abundances of the CN-strong stars this way would require the incorporation of a 
relatively large, and perhaps implausible, amount of ejecta. For example, in order 
to arrive at a 0.8 $M_{\odot}$ star having an abundance of [Fe/H] = $-0.8$,
[C/Fe] = $-0.3$, and [N/Fe] = $+1.4$, by starting with gas that initially
had [C/Fe] = 0.0 and [N/Fe] = $+0.4$, almost half of the stellar mass (0.4
$M_{\odot}$) must come from material essentially devoid of C and extraordinarily
N-rich ([N/Fe] = $+1.7$). The difficulties for a primordial enrichment scenario
presented by such a requirement have been discussed by Smith \& Norris (1982b). 
They show that the large amount of C-depleted material which must be incorporated 
into a population of CN-strong giants imposes rather extreme constraints on the 
mass function of the stars that produced this C-poor material.

One way of trying to circumvent this problem is to assume that the original composition of
a cluster is not represented by the CN-weak stars, but by the C abundances of
the CN-strong stars and the N abundances of the CN-weak stars. From such an
initial mix, enrichment of part of a cluster's stars in N produced CN-strong
stars, while the remaining part of the cluster was enriched in C to
generate the CN-weak stars. The CN-strong stars would have the stronger CN bands
because the percentage increase in N in these stars was greater than the
percentage increase in C in the CN-weak stars. This is an extreme scenario,
requiring that all stars in a cluster have been enriched in one form or
other in the CN(O) elements.

We further note the large N abundance differences implied by the present
analysis are difficult to explain without resorting to a somewhat flat mass
function in either of the above scenarios.
The problem can see be seen if one assumes the cluster formed with its
members homogeneous in their N abundances.
If the low-mass stars were then enriched throughout by N-rich material
ejected from more massive members, how much mass would be required?
Following a Salpeter initial mass function ($dN = m^{-\alpha} dm,$ 
$\alpha=2.35$)
and stars forming within a mass range of 50 to 0.2$M_\odot$, one finds some 50\% of
the total initial cluster mass ($M_t$) will be tied up in stars 1$M_\odot$ or less.
The subsequent ``enrichment'' of half of these low mass stars
with enough N to raise [N/Fe] from 0.4 to 1.4 requires $0.0091M_t$ of nitrogen to be
returned by higher-mass stars (assuming a solar $^{14}$N mass fraction of
$1.01\times10^{-3}$). But the often suggested candidate for such nitrogen production, 
10 - 5$M_\odot$ AGB stars, only account for 8.2\% of $M_t$.
Thus, each AGB star must return a large fraction of its mass
(1.1\%) as N to ``enrich'' the lower-mass stars.

The problem can be ameliorated by considering a less steep mass function
such as is actually observed among main-sequence stars in globular clusters by 
deep HST imaging studies (e.g., Piotto \& Zoccali 1999). 
For example, $\alpha = 1.5$ results in only 8.3\% of $M_t$ in low mass
stars and the need for just $0.00015M_t$ of N to be produced.
With 14\% of the cluster mass in the form of 10 - 5$M_\odot$ AGB stars,
each AGB star need only return a modest 0.11\% of its mass in the form of N.
Indeed, the observations of Hesser et al. (1987) imply a present day
mass function of $\alpha \approx 1.2,$ and De Marchi \& Paresce (1995) found
$\alpha \approx 1.5$ (for $0.8 - 0.3M_\odot$) in 47 Tuc. Similarly, Richer
et al. (1990) derived $\alpha \approx 1.5$ for M71.

Of course the above exercise is entirely speculative. We have neglected the
efficiency with which the low-mass stars incorporate the ejecta,
the source of the C abundance variations, and most notably the likelihood that
these clusters have significantly modified their mass functions through mass
segregation and tidal stripping. 
In addition, lower-mass AGB stars of 2 to 3 $M_\odot$
might also have contributed to cluster enrichment. 
Likewise, the results are sensitive to the
present abundance analysis. A downward revision of the range of [N/Fe] to the
average abundances reported among 47 Tuc main sequence and RGB
stars by Cannon et al. (1998) (i.e., [N/Fe] =0.2 and
1.02 for the CN-weak/strong groups respectively) would require a return of
0.47\% by mass of each AGB star as 
nitrogen with a Salpeter initial mass function.

4) Another constraint on the origin of abundance inhomogeneities within M71 and
47 Tuc is the observation that the range and distribution of CN and CH in these
two clusters are similar, yet M71 has 1/10 the mass of 47 Tuc ($\log
M/M_{\odot} = 4.98$ versus 6.06, as given by Chernoff \& Djorgovski 1989) and a 
very different central concentration ($\log r_t/r_c$ equals 1.15 and 2.04 for M71 
and 47 Tuc, respectively, Harris 1996). Either the mechanism responsible operated 
over a wide range of conditions (which is perhaps difficult to understand in the 
context of the capture of stellar outflows because the central escape speed from 
M71 is only 11 km/s), or M71 was considerably more massive earlier in its history.

In terms of an early enrichment scenario, the acquisition of nuclear-processed material by a 
low-mass cluster star from a former more-massive binary companion might be in
accord with these constraints, as long as M71 and 47 Tuc had similar incidences
of binary stars. In this case, the dynamics of the binary system might had more
of an influence on the enrichment process, by determining the geometry of mass
transfer, than did the cluster environment. In order to avoid having all of the
CN-strong stars still in binary systems, it would be necessary
to invoke that most of these early binaries were eventually disrupted.

\acknowledgements

We wish to express our thanks to Roger Bell whose SSG code was instrumental
in this project and to the Kitt Peak Staff for their assistance in obtaining
these observations. The authors also thank Russell Cannon for a very instructive
referee's report. Partial support was provided by a Theodore Dunham, Jr.
grant for Astronomical Research, and the National Science Foundation under
grants AST-9624680 and AST-0098489 to MMB.

\clearpage

\begin{deluxetable}{ccccccccc}

\tablewidth{0pt}
\tablenum{1}
\tablecaption{Model Parameters and Synthetic Colors for M71 RGB. \label{tbl-1}}
\tablehead{
\colhead{T$_{eff}$}                   &
\colhead{log g}                       &
\colhead{$M_V$}                       &
\colhead{$(B-V)_0$}                   &
\colhead{$(V-K)_0$}                   &
\colhead{$C(41-42)$\tablenotemark{a}}    &
\colhead{$C(42-45)$\tablenotemark{a}}    &
\colhead{$C(41-42)$\tablenotemark{b}}    &
\colhead{$C(42-45)$\tablenotemark{b}}
}
\startdata
3750 & 0.65 & $-$1.50 & 1.507 & 3.94 & 0.183 & 1.281 & 0.313 & 1.254\nl
4000 & 1.05 & $-$1.12 & 1.276 & 3.41 & 0.174 & 1.112 & 0.352 & 1.070\nl
4250 & 1.55 & $-$0.37 & 1.092 & 3.01 & 0.129 & 0.989 & 0.332 & 0.942\nl
4500 & 2.05 &   +0.47 & 0.950 & 2.67 & 0.088 & 0.902 & 0.287 & 0.853\nl
4750 & 2.55 &   +1.36 & 0.835 & 2.39 & 0.059 & 0.833 & 0.232 & 0.785\nl
5000 & 3.10 &   +2.43 & 0.737 & 2.15 & 0.039 & 0.777 & 0.170 & 0.730\nl
\enddata

\tablenotetext{a}{[C/Fe] = 0.0, [N/Fe] = +0.4, [O/Fe] = +0.4}
\tablenotetext{b}{[C/Fe] = $-0.3$, [N/Fe] = +1.4, [O/Fe] = +0.2}
\end{deluxetable}

\clearpage

\begin{deluxetable}{ccccccccc}

\tablewidth{0pt}
\tablenum{2}
\tablecaption{Synthetic Colors for M71 RGB with Deep Mixing. \label{tbl-2}}
\tablehead{
\colhead{$M_V$}        &
\colhead{[C/Fe]\tablenotemark{a}}      &
\colhead{[N/Fe]\tablenotemark{a}}     &
\colhead{$C(41-42)$\tablenotemark{a}}    &
\colhead{$C(42-45)$\tablenotemark{a}}    &
\colhead{[C/Fe]\tablenotemark{b}}      &
\colhead{[N/Fe]\tablenotemark{b}}     &
\colhead{$C(41-42)$\tablenotemark{b}}    &
\colhead{$C(42-45)$\tablenotemark{b}}
}
\startdata
$-$1.50 & $-1.00$ & +0.73 & 0.167 & 1.209 & $-1.00$ & +1.42 & 0.216 & 1.218 \nl
$-$1.12 & $-0.75$ & +0.71 & 0.167 & 1.028 & $-0.84$ & +1.42 & 0.243 & 1.027 \nl
$-$0.37 & $-0.50$ & +0.67 & 0.135 & 0.914 & $-0.67$ & +1.42 & 0.240 & 0.903 \nl
  +0.47 & $-0.25$ & +0.59 & 0.102 & 0.855 & $-0.50$ & +1.41 & 0.234 & 0.828 \nl
\enddata
\tablenotetext{a}{CN-weak model with mixing}
\tablenotetext{b}{CN-strong model with mixing}
\end{deluxetable}

\clearpage

\begin{figure}
\epsscale{0.7}
\plotone{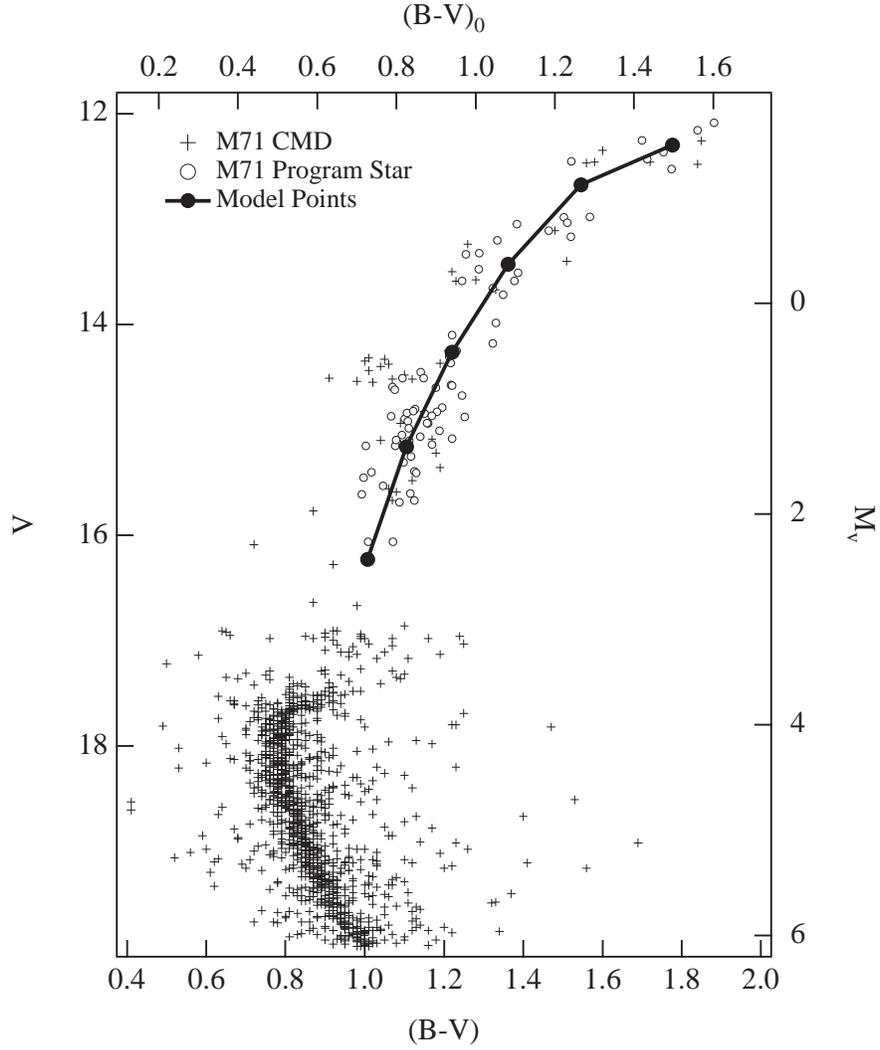}
\caption[Briley.fig1.eps]{The observed color-magnitude diagram of M71 from the
photometry of Hodder et~al. (1992). Filled circles indicate the locations of
the model points from Table 1. The colors and brightnesses of the 75 RGB stars with 
DDO photometry are plotted with open circles (using the KPNO data.
\label{fig1}}
\end{figure}

\begin{figure}
\epsscale{0.7}
\plotone{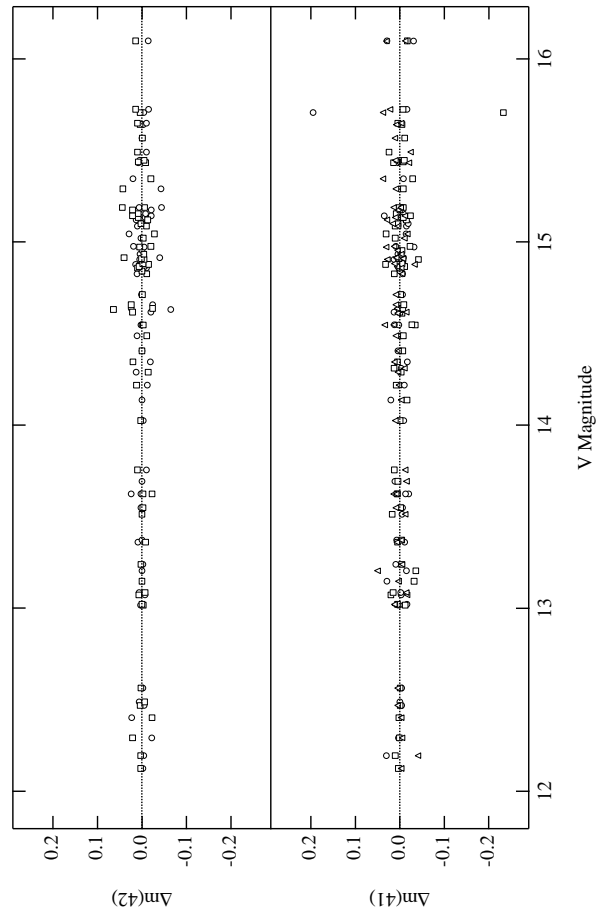}
\caption[Briley.fig2.eps]{Photometric errors in the 41 and 42 magnitudes determined from 
multiple exposures are plotted as a function of brightness for the program stars.
\label{fig2}}
\end{figure}

\begin{figure}
\epsscale{0.7}
\plotone{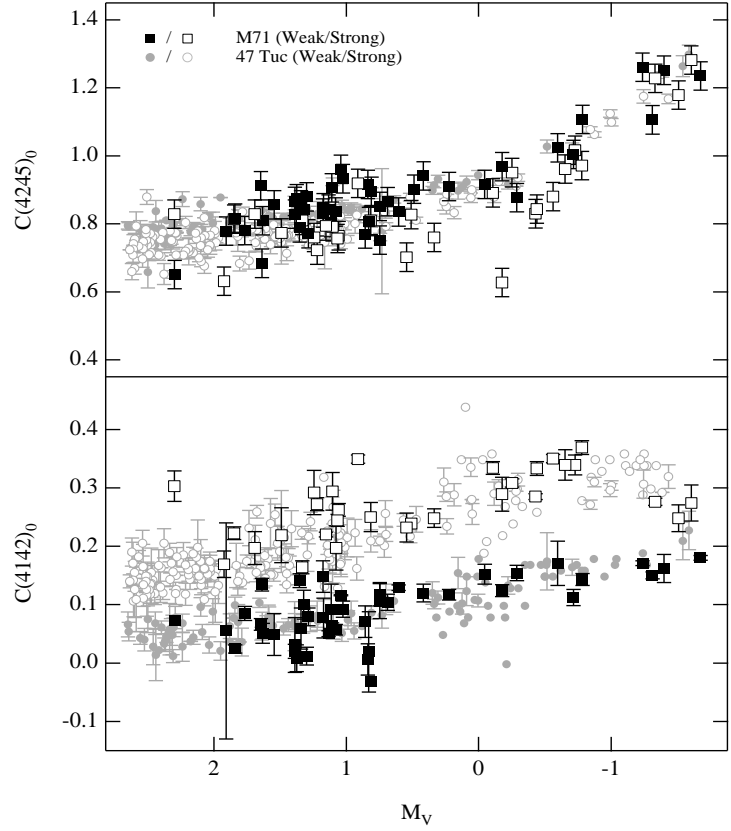}
\caption[Briley.fig3.eps]{Reddening corrected DDO $C(41-42)$ and $C(42-45)$ colors for 
75 M71 red giants (square symbols). Also plotted are the DDO colors for 47 Tuc 
giants from Norris \& Freeman (1979) and Briley (1997) (circles).
\label{fig3}}
\end{figure}

\begin{figure}
\epsscale{0.7}
\plotone{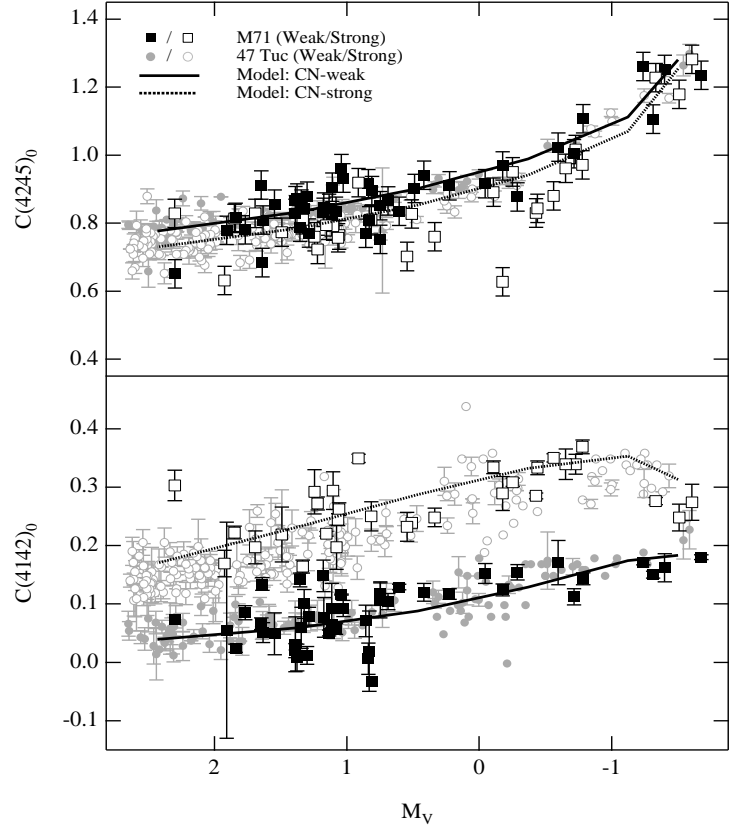}
\caption[Briley.fig4.eps]{The photometry from Figure 3 is plotted with the synthetic 
colors from Table 1. The model colors correspond to constant abundances of
[C/Fe] = 0.0, [N/Fe] = +0.4, [O/Fe] = +0.4 (for the CN-weak stars)
and [C/Fe] = $-0.3$, [N/Fe] = +1.4, [O/Fe] = +0.2 (for the CN-strong). These
same abundance combinations were found by Briley \& Cohen (2001) to match the CN
and CH band strengths of M71 main-sequence stars.
\label{fig4}}
\end{figure}

\begin{figure}
\epsscale{0.7}
\plotone{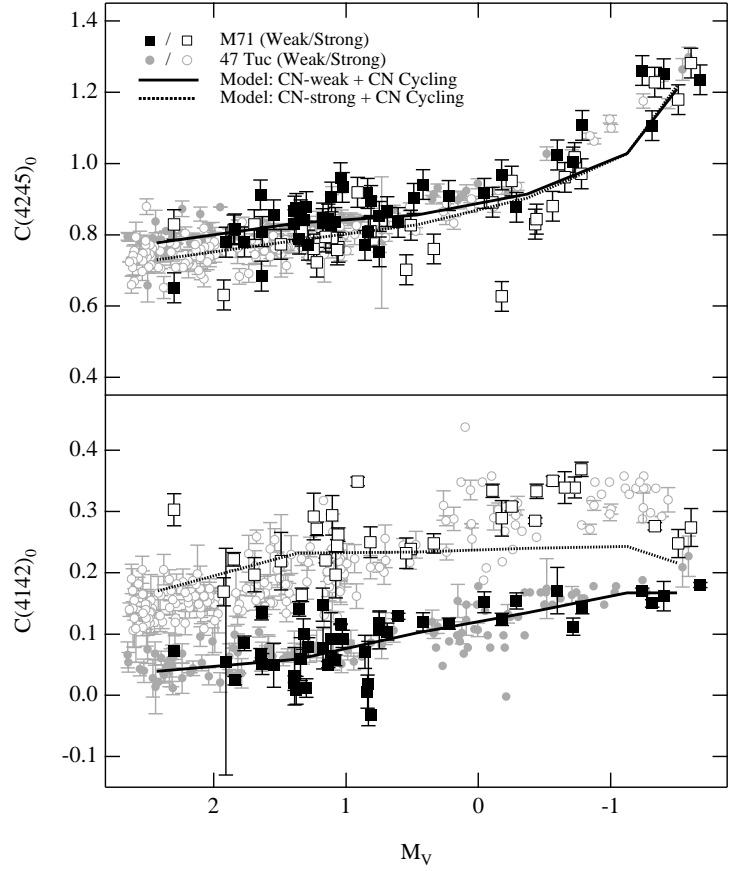}
\caption[Briley.fig5.eps]{The synthetic colors of Table 2, which correspond to the 
expected result of an extreme mixing of CN-cycled material to the stellar surface 
during RGB ascent, are plotted with the observed DDO colors.
\label{fig5}}
\end{figure}

\end{document}